\documentclass[prl,showkeys,twocolumn]{revtex4}
\usepackage{graphicx,amssymb,amsmath}
\usepackage{hyperref}
\usepackage{enumitem}
\usepackage{empheq}
\usepackage{color,soul}
\usepackage{mathrsfs}
\usepackage{afterpage}
\usepackage[caption=false]{subfig}
\usepackage{url}

\begin{document}

\title{Stochastic equations and cities}

\author{Marc Barthelemy}
\email{marc.barthelemy@ipht.fr}
\affiliation{Universit\'e Paris-Saclay, CEA, CNRS, Institut de Physique Th\'eorique, 91191, Gif-sur-Yvette, France}
\affiliation{Centre d'Analyse et de Math\'ematique Sociales, (CNRS/EHESS) 54, Boulevard Raspail, 75006 Paris, France}

\begin{abstract}

Stochastic equations constitute a major ingredient in many branches of science, from physics to biology and engineering. Not surprisingly, they appear in many quantitative studies of complex systems. In particular, this type of equation is useful for understanding the dynamics of urban population. Empirically, the population of cities follows a seemingly universal law - called Zipf's law - which was discovered about a century ago and states that when sorted in decreasing order, the population of a city varies as the inverse of its rank. Recent data however showed that this law is only approximate and in some cases not even verified. In addition, the ranks of cities follow a turbulent dynamics: some cities rise while other fall and disappear. Both these aspects - Zipf's law (and deviations around it), and the turbulent dynamics of ranks - need to be explained by the same theoretical
framework and it is natural to look for the equation that governs the evolution of urban populations. We will review here the main theoretical attempts based on stochastic equations to describe these empirical facts. We start with the simple Gibrat model that introduces random growth rates, and we will then discuss the Gabaix model that adds friction for allowing the existence of a stationary distribution. Concerning the dynamics of ranks, we will discuss a phenomenological stochastic equation that describes rank variations in many systems - including cities -  and displays a noise-induced transition. We then illustrate the importance of exchanges between the constituents of the system with the diffusion with noise equation. We will explicit this in the case of cities where a stochastic equation for populations can be derived from first principles and confirms the crucial importance of inter-urban migrations shocks for explaining the statistics and the dynamics of the population of cities.

\end{abstract}

\maketitle

\section{Introduction}

Even if there is no universally accepted definition of complex systems, cities can certainly be considered as one of their most prominent example \cite{Batty:2013,Barthelemy:2016,Bettencourt:2021}. Cities are made of a large number of different constituents interacting on multiple spatial and temporal scales, leading to the emergence of multiple collective behaviors such as traffic jams, gentrification, etc. Obviously, there are many aspects in cities and a single`unified' model describing all of them seems out of reach for the moment. There are however simple quantities that describe the importance of a city and help monitoring their evolution. Among these quantities such as the surface area, the GDP, energy consumption, etc., the population appears as fundamental. Knowing this quantity about a city gives informs us about its structure \cite{Pumain:2004}, and when combined with its location about its importance in the country \cite{Louail:2022}. 

The population of a city depends on the definition of its frontiers. Administrative boundaries are certainly now outdated and it is now more accurate to speak about urban areas. Standard definitions used now involve the identification of a urban center and connect this center to surrounding areas exchanging a number of commuters large enough (the so-called Metropolitan Statistical Areas in the US and the Larger Urban Zones - LUZ - in Europe). Another interesting definition relies on percolation and defines the city as the giant component of built-areas \cite{Rozen:2011}. It is important to be aware of this problem, but we won't focus on this here, and we will mostly consider data for the population of urban areas defined in standard ways. 

The population of cities displays a very large range of variation, from small towns with about $10^4$ inhabitants to megacities with a population of about $10^7$ individuals. This broadness rules out any type of optimum argument where the concentration of individuals would lead to some cost-benefit equilibrium implying a unique typical size of cities. This sort of optimum discussion was very common in economics \cite{Henderson:1974}, but seems in contradiction with empirical facts. Indeed, the statistics 
of urban population follows a universal law \-- called Zipf's law \cite{Zipf}   \-- and was
discovered almost a century ago by the physicist Auerbach \cite{Auerbach}. Auerbach sorted the population $P$ of german cities in decreasing order (the rank $r=1$ corresponds then to the largest city, Berlin at that time), and noticed that the product $P\times r$ was roughly constant. Equivalently, Zipf's law for cities (a law extended by Zipf to many other systems) states that the population $P$ of a city varies approximately as
the inverse of its rank $r$. This empirical discovery led to almost a century of theoretical studies (see \cite{Arshad} for a recent review) mostly in economics \cite{Singer, Krugman, Ioannides, Eeckhout, Rossi, Cordoba, Favaro, Duranton, Gabaix}, in geography \cite{Pumain} and in physics  \cite{Blank,Coro,Benguigui,Zanette,Marsili}. The recent availability of data for different countries and time periods brought however new facts.

First, the validity of Zipf's law has been challenged \cite{Soo, Gan, Benguigui2, Cottineau,Arshad}.  In many cases, the Zipf plot is not displaying a simple $1/r$ behavior \cite{Arshad} and the distribution of city size can appear to be much more complex as a power law with candidate functions such as the lognormal or the stretched exponential \cite{Sornette1,Arshad}.

Second, the ranks of cities follow a turbulent dynamics with sometimes very large fluctuations \cite{Batty:2006}: some cities rise while other fall and disappear. Both these aspects - Zipf's law and the turbulent dynamics - are obviously connected and need to be explained in the same theoretical framework. We will describe here the various  attempts to describe these
empirical facts, and that are all based on stochastic equations governing the
time evolution of urban populations. 

A stochastic differential equation (SDE) is a differential equation  where one or more terms or coefficients are random (usually white noise but as we will see the noise can be more complex). These random terms can be constants or functions and their statistical properties are generally supposed to be given. As a result, the solution is also a stochastic process and can be described by its distribution or by its first moments \cite{VanKampen,Bernt}.  These ubiquitous equations were discussed at length in the mathematical literature and are naturally found in physics, but also in many other fields ranging from ecology to finance. The random terms in these equations usually describes the effect of the interaction of many elements for which it is impossible to describe all the details \cite{VanKampen}.  The broad and interdisciplinary field of complex systems  concerns essentially studies of individual entities that interact and produce collective dynamics. It is therefore not a surprise that stochastic equations appear in this context as we will verify it for cities. 

In the first part of this article, we describe the most salient empirical features about urban populations: the distribution of population and Zipf's law,  and the dynamics of city ranks. We will then discuss recent empirical results that demonstrate the existence of important deviations from the Zipf law that was thought to be universal. In a second part, we introduce the historical models for understanding Zipf's law: the Gibrat model of random growth and its limits, and then Gabaix' model which is equivalent
to a random walk with a reflecting barrier and which allows to understand certain aspects of the population distribution. We note here that other models were proposed for explaining the emergence of a power law or other functions and go from simple stochastic processes to economic models with a large number of parameters. In \cite{Benguigui}, the authors distinguish models with a constant number of cities and redistribution between these cities, while other models propose mechanisms for city growth and a varying number of cities (\cite{Benguigui} and references therein). Here, we focus on the Gabaix model as it is so far commonly considered as the natural explanation for Zipf's law. As a digression, we will describe a general phenomenological stochastic equation that allows to understand rank fluctuations in various systems and which displays an interesting noise-induced transition. In the last part of this paper, we will see that these models (Gibrat, Gabaix) are not able to explain the turbulent dynamics of cities or the observed deviations from Zipf's law. This will lead us to review the diffusion with noise equation discussed in another context (the condensation of wealth in a simple  model of economy). In the city context, this suggests the critical importance of interurban migration flows for understanding the population statistics. Indeed, as we will show in the last part of the paper, by starting from first principles we can derive an equation similar to the diffusion with noise and that describes the growth of cities. The derivation combines empirical arguments, and the generalized central limit theorem. This stochastic equation with multiplicative noises predicts a complex shape for the distribution of city populations and shows that, owing to finite-time effects, Zipf’s law does not hold in general, implying a more complex organization of cities. It also predicts the existence of multiple temporal variations in the
city hierarchy, in agreement with observations. We will then end this paper with a general discussion and review some of the current challenges in this field.

\section{Empirical studies}

\subsection{Rank-size plot and population distribution}

The first observation of a statistical regularity in the urban population distribution was
made in 1913 by Felix Auerbach \cite{Auerbach}, a german physicist interested in interdisciplinary studies. Auerbach first ranks in decreasing order cities of Germany (rank $1$ is the largest city, Berlin at that time, and the last rank $N$ corresponds to the smallest city in the study). A given city has then a rank $r$ and a population $P$, and Auerbach noticed that the product of these quantities is approximately constant
\begin{align}
	r\times P\approx \mathrm{const.}
	\label{eq:auer}
\end{align}
where the constant for Germany was at that time is $\approx 50$. This first observation was generalized by Zipf \cite{Zipf} who ranked many items (such as the number of times a word appears in a book) and proposed to plot the score of an item versus its rank. This rank plot - also called Zipf plot now- became a useful tool for the analysis of many complex systems. 

\begin{figure}[!h]
	\centering
	\includegraphics[scale=0.55]{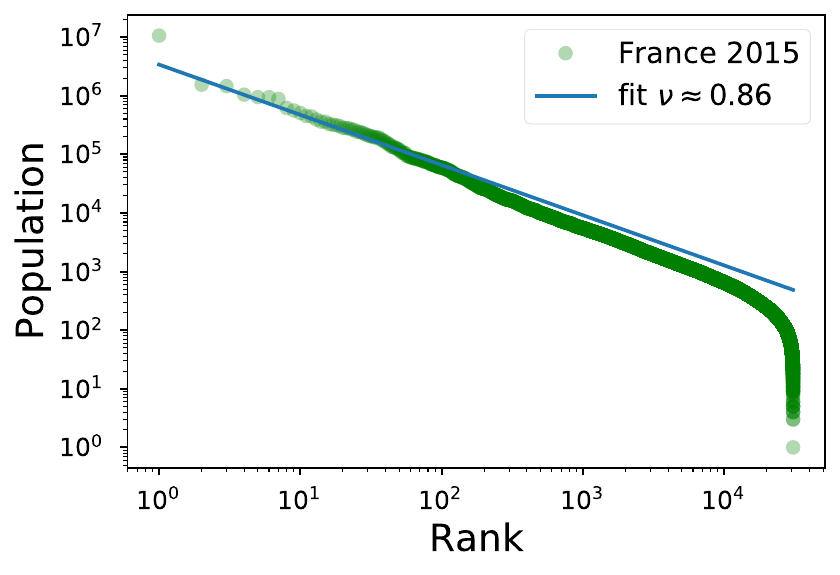}
	\caption{Zipf (or rank) plot for cities in France for the year 2015. The power law fit over the rank range $[2,10^4]$ gives an exponent $\nu\approx 0.86$. }
	\label{fig:zipf}
\end{figure}

This relation Eq.~\ref{eq:auer} means that the rank $r$ scales as $1/P$. In general, we observe the more general power law
\begin{align}
	P\sim \frac{1}{r^\nu}
	\label{eq:zipf}
\end{align}
where $\nu$ was found initially approximately equal to $1$ by Auerbach and Zipf \cite{Zipf}, but was found in subsequent studies to take various values \cite{Soo,Cottineau} (see Fig.~\ref{fig:zipf} for an example of a Zipf plot with $\nu\neq 1$). 

This Zipf relation Eq.~\ref{eq:zipf} is all but trivial and has many consequences that 
were discussed many times in the literature (see for example
\cite{Pumain:2004} and references therein). It implies that for a given system (a country for example), if we know the population of a city (for example the largest one $P_1$), we can then determine the values of all the others using $P(r)=P_1/r^\nu$. The second largest city is then the half of the largest one, etc. In addition, if we denote by $\rho(P)$ the distribution of the population, the rank is then approximately given by
\begin{align}
	r(P)\sim N\int_P^{P_{\max}}\rho(P)\mathrm{d}P
\end{align}
Indeed, if $P=P_{\min}$, we obtain $r(P)=N$, and if $P=P_{\max}$, the integral is of order $1/N$ and $r(P_{\max})=1$.  From this expression, we then obtain for $P_{\max}\gg P$ that the distribution is also a power law 
\begin{align}
	\rho(P)\sim \frac{1}{P^\mu}
	\label{eq:rhoP}
\end{align}
with $\mu=1+1/\nu$. For the value $\nu\approx 1$ initially observed by Auerbach, this relation implies that the population distribution has a fat tail decaying as $\rho(P)\sim 1/P^2$. 

The Zipf law also allows to estimate how the size of the largest city or the number of cities in a country vary with the total population. Indeed, if we denote by $N$ the number of cities, by $W$ the total urban population in the country, we have the following relation
\begin{equation}
	W=\sum_{i=1}^NP_i
	\label{eq:sum}
\end{equation}
The exponent $\mu$ in Eq.~(\ref{eq:rhoP}) is around $2$ and precise
values are either below or above. When the exponent is $\mu>2$, the
dominant behavior of Eq.~(\ref{eq:sum}) is given by the usual central
limit theorem
\begin{equation}
	W\sim N
\end{equation}
and the number of cities is thus proportional to $W$. When the
exponent is smaller than $2$, we have a sum of broadly distributed
variables with divergent average and the sum is dominated by the
largest and behaves as (see for example \cite{Bouchaud:1990})
\begin{equation}
	W\sim N^{1/(\mu-1)}
\end{equation}
which implies that the number of cities scales as $N\sim W^{\mu-1}$.
We thus see that in both cases the number of cities varies as
\begin{equation}
	N\sim W^\zeta
\end{equation}
where $\zeta$ is either exactly or numerically close to one.

\begin{figure}[!ht]
\centering
\includegraphics[scale=0.5]{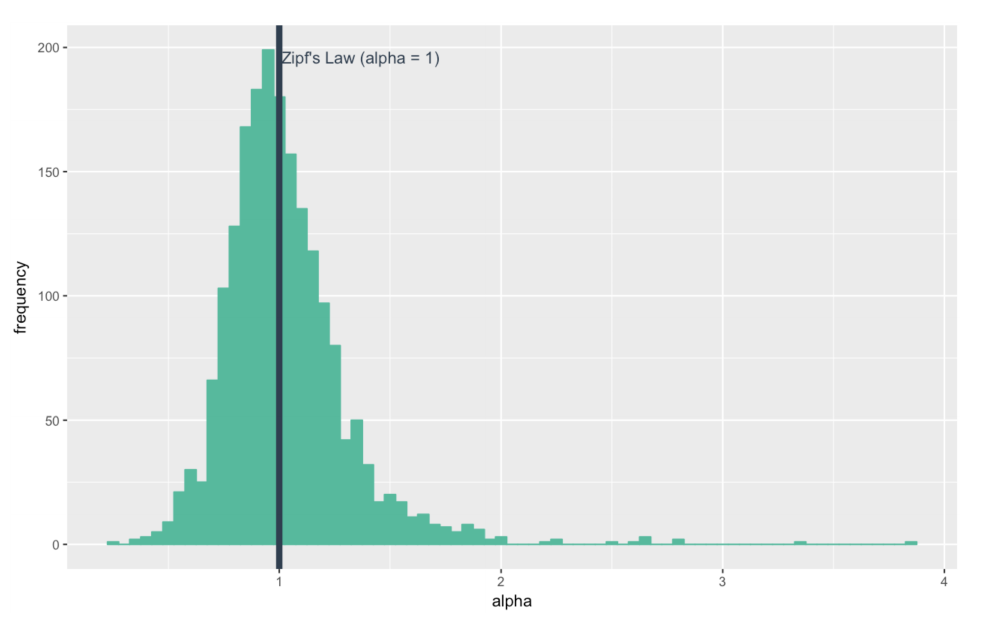}
 \caption{Diversity of the Zipf exponent (data for almost 2000 values compiled by \cite{Cottineau}). Figure from \cite{Cottineau}.}
\label{fig:cotti}
\end{figure}

Most of the empirical measures used the generalized form Eq.~\ref{eq:zipf} and determined the so-called Zipf exponent $\nu$. The original works of Auerbach and Zipf discussed essentially the case $\nu=1$.  Many measures were done since then, for various countries, different epochs, and most of these results were compiled in the study \cite{Cottineau}. The resulting distribution of the Zipf exponent $\nu$ is shown in Fig.~\ref{fig:cotti}.
We see in this figure that even if the value $\nu=1$ is indeed the most probable, there is a non negligible dispersion around this value. In addition, it must be noted that the Zipf plot is not always a clear power law and can display an important curvature (in loglog) signaling the emergence of more complex functions that just a simple power law. All these empirical results need to be explained and this will be the goal of the theoretical analysis presented below.

\subsection{Dynamics of ranks: rises and falls of cities}

\begin{figure}
	\centering
	\includegraphics[scale=0.4]{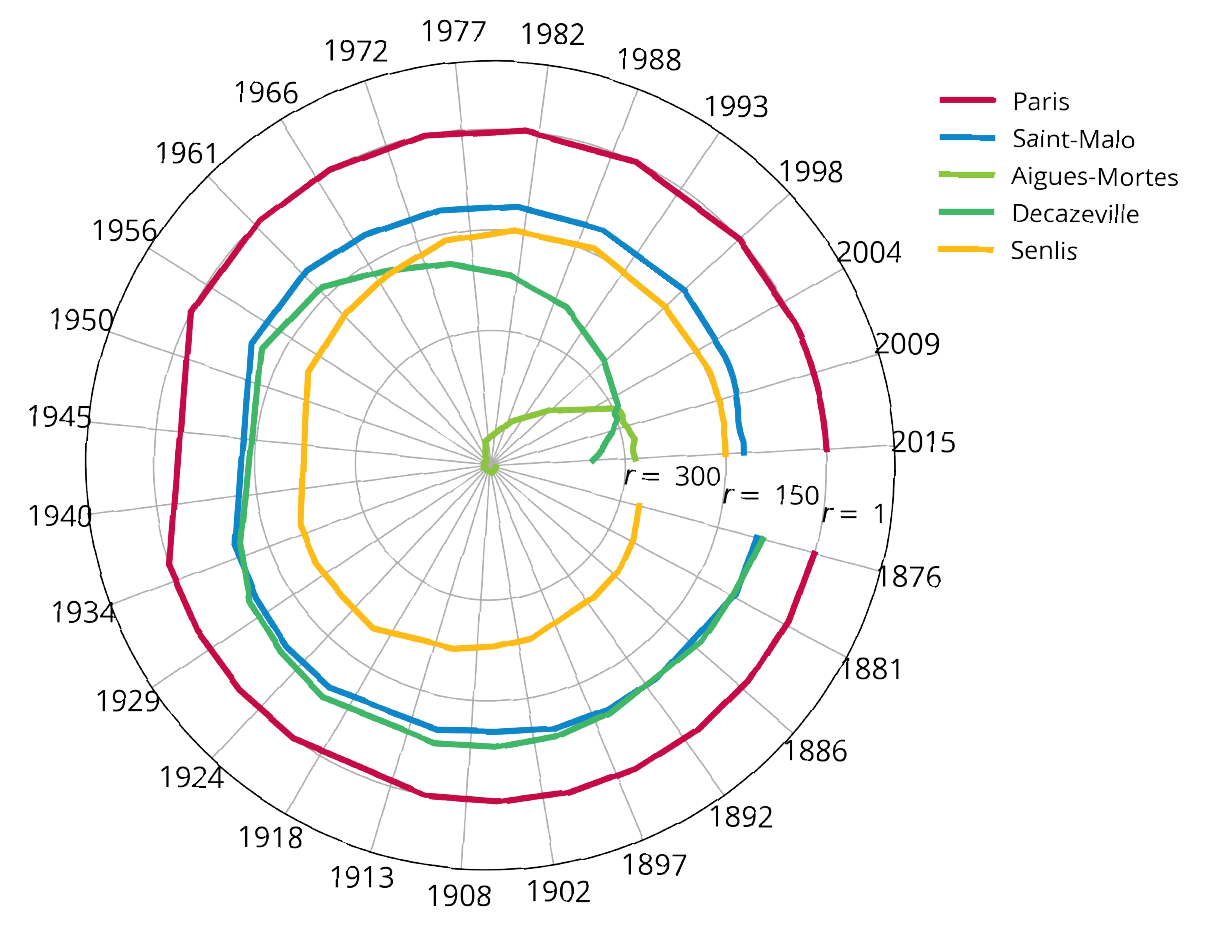}
	\caption{Rank clock for a few selected cities in France. The first rank is shown here at the external border of the clock, and the angle is proportional to the time. Figure courtesy by V. Verbavatz.}
	\label{fig:FRrankclock}
\end{figure}

In addition to the existence of an apparently stationary
distribution, the ranks of cities follow a turbulent dynamics: some cities
rise while other fall and disappear. The dynamics of populations and their rank in the country can be visualized with the help of `rank clocks' proposed in \cite{Batty:2006}. In these rank clocks, the radius is proportional to the rank (rank $1$ has the largest radius) and the angle is proportional to time. We can then follow the trajectory of a city during time and see how its rank evolves. We show an example (Fig.~\ref{fig:FRrankclock}) of such a rank clock for few cities in France. 

In \cite{Batty:2006}, the geographer M. Batty showed that the micro-dynamics of cities is very turbulent and displays rises and falls of entire cities at many times
and on many scales. As we will see in the next sections, this puts important constraints on possible models. The dynamics cannot be smooth but somehow must explain the appearance of shocks that can significantly modify the ranks of certain cities.

\section{Explaining Zipf's law with stochastic equations}

Most of the first studies on Zipf's law focused on the case $\nu=1$ where the rank $r$ and the population $P$ are inversely proportional $r\sim 1/P$. We will here discuss one of the most important historical model proposed by Gabaix \cite{Gabaix} which is able to explain this value for the Zipf exponent. We will first discuss the simplest stochastic model for describing the random growth of a city:  the Gibrat model.

\subsection{The Gibrat model (proportionate growth)}

Proposed originally by Gibrat in 1931 \cite{Gibrat} for explaining the growth of firms, this model 
assumes a random growth where the rate is a random variable, independent from the size of the firm.  Transposed to the population $P_i$ of city $i$, this reads as
\begin{align}
	P_i(t+1)=\eta_i(t)P_i(t)
	\label{eq:gibrat}
\end{align}
where the growth rates $\eta_i(t)$ are random positive, independent variables, and distributed according to the same distribution $f(\eta)$.  We assume here that time is discretized and typically $t$ represents a year. If this relation is always valid, by taking the logarithm we obtain
\begin{align}
	\ln P_i(t)=\ln P_i(0)+\sum_{\tau=0}^{t-1}\ln\eta_i(\tau)
\end{align}
and according to the central limit theorem (see for example \cite{Fisher}), the variable $\ln P$ is Gaussian, leading to a lognormal distribution as the variance $var(\ln P)\sim var(\ln\eta)t$ and diverges at large times. In this case, the absence of a steady state leads to a lognormal distribution in disagreement with empirical observations. In order to reconciliate a random growth and the empirical observation of Zipf, it is necessary to introduce some sort of friction that implies the existence of a steady state. This is the main point of the Gabaix model discussed in the next section. 

\subsection{The Gabaix model (random walk with a reflecting barrier)}


In order to understand the Zipf law, Gabaix \cite{Gabaix} proposed a variant of the Gibrat model based on the idea that small cities cannot shrink to zero. In other words, the process considered by Gabaix is a random walk with a lower reflecting barrier which can, in some conditions, produces a power law distribution of populations. There are two essential assumptions here. The first ingredient is the existence of a perfectly reflecting barrier which prevent cities from becoming too small. The second ingredient is that $\langle\ln\eta\rangle<0$ (the brackets denote here and in the following the average over the distribution of $\eta$). Indeed, if it is not the case, the walk will escape to infinity and we recover the central limit theorem and a lognormal distribution for $P$. Instead, if the walker is pushed towards zero ( $\langle\ln\eta\rangle<0$), it will bounce on the reflecting barrier and create an effective flow going to large values. This interplay between a drift towards the barrier and the reflection at the barrier is what gives rise to a power law.

This problem of a multiplicative noise with a reflecting barrier was solved by Levy and Solomon \cite{Levy} and discussed in more detail by Sornette and Cont \cite{Sornette}. We first introduce $w=P/\overline{P}$ where $\overline{P}$ is the average population. If we assume that $d\ln \overline{P}/dt$ is negligible compared to the average growth rate, we obtain for $w$ an equation of the form
\begin{align}
	w(t+1)=\eta (t)w(t)
\end{align}
where we redefined the average of $\eta$. We change variables $x = \ln w$ and $\ell =\ln\eta$ and assume that the reflecting wall is at $x_0 = \ln w_0$. In this process, the position of the walker at time $t + 1$ is given by (we also assume here that time is discretized)
\begin{align}
	x(t+1)=\max (x_0,x(t)+\ell)
\end{align}
where $\ell$ is the value of the increment (and distributed according to $\pi(\ell)$). This equation means that $x_{t+1} = x_0$ if the jump brings the walker below the barrier, and $x_{t+1} = x_t + \ell$ is the new position if it is above the barrier. The walker will then be at $x_0$ with probability
\begin{align}
	P(x=x_0)=\int_{-\infty}^{x_0}\mathrm{d}x\int_{-\infty}^{+\infty}P(x-\ell)\pi(\ell)\mathrm{d}\ell
\end{align}
where $P(x-\ell)$ is the probability to be at position $x-\ell$ and $\pi(\ell)$ is the probability to jump a distance $\ell$. For $x>x_0$, we have
\begin{align}
	P(x)=\int_{-\infty}^{+\infty}P(x-\ell)\pi(\ell)\mathrm{d}\ell
	\label{eq:self}
\end{align}
This is a Wiener-Hopf equation and solving it can be mathematically involved. Here, we proceed in a simplified way and assume an exponential ansatz $\mathrm{e}^{\gamma x}$ for the solution. Plugging this form into Eq.~\ref{eq:self} leads to the equation that $\gamma$ must satisfy
\begin{align}
1=\int \mathrm{e}^{\gamma \ell} \pi(\ell)\mathrm{d}\ell
\end{align}
The stationary limit is then $P(x) = \mathrm{e}^{\gamma x}$ which implies that the
distribution of $w$ is given by a power law of the form
\begin{align}
	P(w)\sim \frac{C}{w^{1+\gamma}}
\end{align}
for $w>w_0$. This result was discussed by Sornette and Cont \cite{Sornette} who proposed another derivation for it. In particular, they proposed the following intuitive explanation for the appearance of a power law: in the presence of a reflecting barrier which prevents the random walk to escape towards very small values, there is a continuous flow of particles that can sample the large positive values of $x$, leading to a broad tail for $P(w)$.

If we now write the two conditions $\int_w P(w)=1$ and $\langle w\rangle =1$, we obtain the relation 
\begin{align}
	\gamma = \frac{1}{1+w_0}
\end{align}
which shows that if $w_0$ is small enough, we have $\gamma \approx 1$. This framework thus justifies the Zipf law (and equivalently a power distribution for population) with an exponent $\nu=\gamma\approx 1$. The price to pay is to introduce a minimal population and that  $\langle \ln\eta\rangle <0$. These assumptions are a bit difficult to justify. In addition, as we will see in the part about the dynamics, even if the Gabaix model is able to explain some of the empirical features about the population distribution, it is not able to explain dynamical features such as large rank variations.

We note here that this model is equivalent to the process discussed by Kesten \cite{Kesten} and which reads
\begin{align}
	\frac{dw}{dt}=\eta(t)w(t)+\epsilon(t)
\end{align}
where $\epsilon(t)>0$ is a positive noise.


\subsection{Digression: A phenomenological stochastic equation for the ranking dynamics}

The problem of the rank dynamics of cities is in fact more general as we can rank approximately anything and as such constitutes an interesting problem with many applications. Indeed, we can consider many examples such as the number of times individual words are used in published journal during a year, the daily market capitalization of companies, the number of diagnoses of a particular disease, the number of annual citations each paper received in the Physical Review corpus, etc. \cite{Blumm,Iniguez}. The dynamics of ranking was discussed phenomenologically in \cite{Blumm} and we will follow this analysis even if it was not designed specifically for cities.  In the ranking problem, we have in general a list of $N$ items with some score $X_i(t)$  which determines their ranking (for cities $X$ is their population). The highest score corresponds then to the first rank $X_i=\max\{X_j\}\Rightarrow r_i=1$ and the smallest score to the last rank $r=N$.  The total number of times and the total score can vary and it is natural 
to consider the normalized score \cite{Blumm}
\begin{align}
	x_i(t)=\frac{X_i(t)}{\sum_jX_j(t)}
\end{align}

\begin{figure}[ht!]
	\centering
	\includegraphics[width=0.9\linewidth]{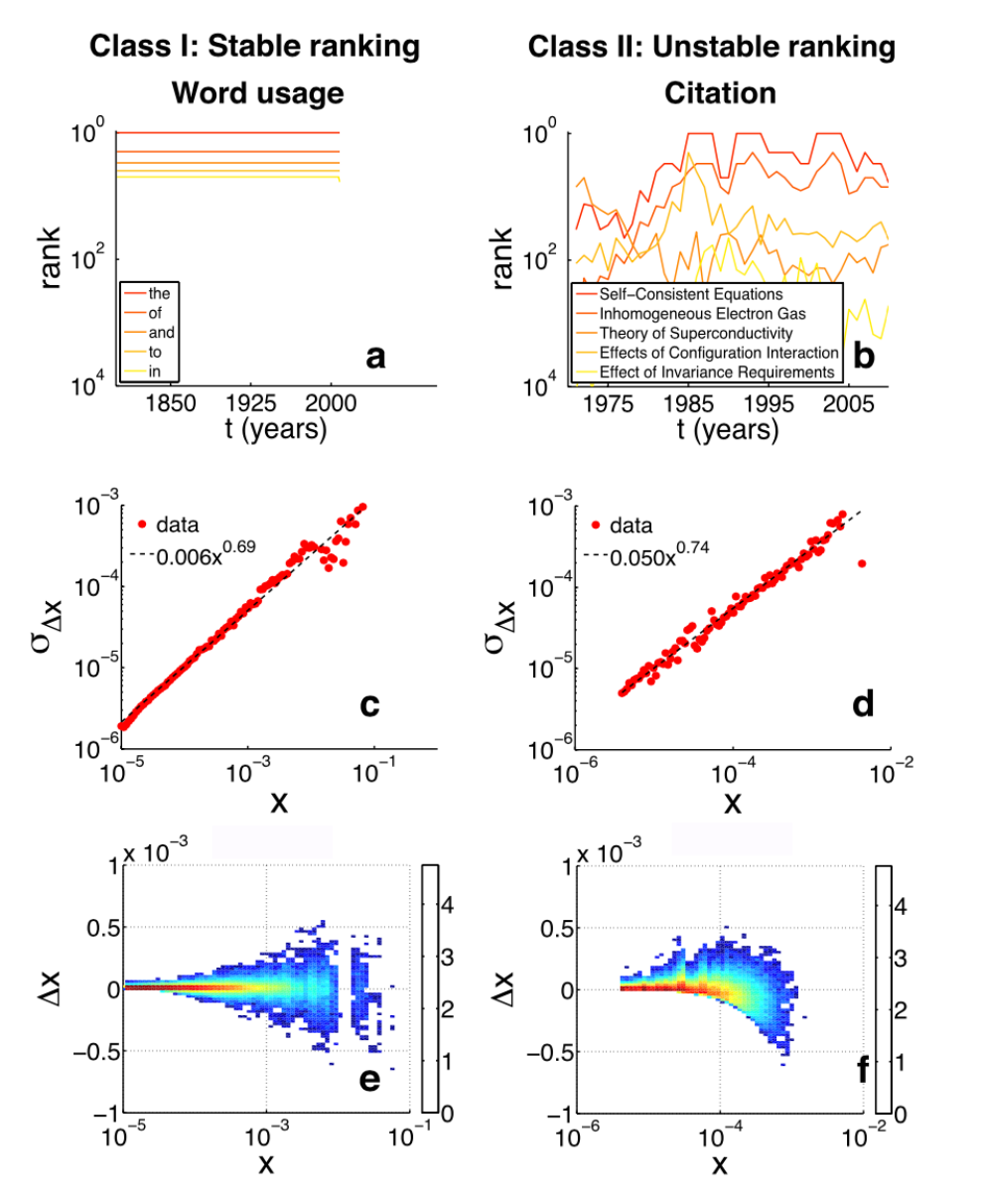}
	\caption{Empirical characteristics of ranking dynamics for word usage (left column) and citations (right column). (Top) On the left, we see an example of stable ranking and on the right column an example of unstable ranking. (Middle) The dispersion of the score varies with its values as a power law. (Bottom) Surface plot in the plane $x,\Delta x$ indicating that for the word usage case the score $x$ fluctuates symmetrically around $0$, while fluctuations are asymmetric for large $x$ in the citation case. Figure from \protect\cite{Blumm}.}
	\label{fig:blumm1}
\end{figure}

In general, the distribution of $x$ follow a broad law spanning many orders of magnitude. In some case, we observe a clear power law behavior, while for other there is cut-off for large $x$ \cite{Blumm}. The dynamics can also be different with cases presenting a stationary law while others display time variations. Despite these differences, there are some `universal' features (see Fig.~\ref{fig:blumm1}). For example, the dispersion of the change $\Delta x$ as a function of $x$ follows the behavior
\begin{align}
	\sigma_{\Delta x|x}\sim x^\beta
\end{align} 
where $\beta$ is in general smaller than $1$. 

In order to discuss the dynamics of ranking, the authors of \cite{Blumm} assumed that
the score of an item $i$ follows a simplified Langevin equation of the form (already discussed in \cite{BouchaudMezard})
\begin{align}
	\frac{dx_i}{dt}=f(x_i)+g(x_i)\xi_i(t)-\phi(t)x_i
	\label{eq:bou1}
\end{align}
where $f(x)$ represents the deterministic process that governs the score and captures many attributes such as the impact of research papers, the utility of a word, etc. The multiplicative noise term of the form $g(x_i)\xi_i(t)$ captures the inherent randomness in the system. The noise $\xi_i(t)$ is assumed to be gaussian with $\langle\xi_i(t)\rangle=0$ and $\langle\xi(t)\xi(t')\rangle=\delta(t-t')$) and the noise amplitude $g(x_i)$ can in general depend on the score $x_i$. Finally, the last term $\phi(t)x_i$ ensures that the scores are normalized $\sum_ix_i(t)=1$. If we sum over $i$ Eq.~\ref{eq:bou1}, we find
\begin{align}
	\nonumber
	\sum_i\dot{x}_i&=0\\
	&=\sum_if(x_i)+\sum_ig(x_i)\xi_i(t)-\phi(t)\sum_ix_i
\end{align}
which implies that
\begin{align}
	\nonumber
	\phi(t)&=\sum_i[f(x_i)+g(x_i)\xi(t)]\\
	&=\phi_0+\eta(t)
\end{align}
where the constant is $\phi_0=\sum_if(x_i)$ and the global noise term is $\eta(t)=\sum_ig(x_i)\xi_i(t)$. 

The equation (\ref{eq:bou1}) is obviously too general and the empirical data suggest some simplifications. First, the drift term $f(x)$ can be written as 
\begin{align}
	f(x_i)=A_ix_i^{\alpha}
\end{align}
where it is assumed that the exponent $0<\alpha<1$ (in order to get a positive solution) is the same for all $i$. The prefactor $A_i$ can be interpreted as the fitness of item $i$ and captures its ability to increase its share $x_i$ (the larger $A_i$ and the larger the growth rate $\dot{x}_i$). The authors of \cite{Blumm} also assume that the noise amplitude behaves as
\begin{align}
	g(x_i)=Bx_i^\beta
\end{align}
which is suggested by the empirical measurements on the variance shown in Fig.~\ref{fig:blumm1}(c,d) (indeed Eq.~\ref{eq:bou1} implies that $\sigma^2_{\Delta x_i|x_i}\approx g(x_i)^2\Delta t$). Empirically, the exponent $\beta$ seems to be comparable for all systems and the amplitude $B$ displays significative differences and varies from $B\approx 10^{-3}$ to $10^{-1}$  \cite{Blumm}.

The coefficient $B$ is a measure of the noise magnitude and we expect that the stability of the system will be affected by its value. Indeed, if we denote by $P(x_i,t|A_i)$, the probability that an item $i$ has score $x_i$ at time $t$ given its fitness $A_i$, its temporal evolution is governed by the Fokker-Planck equation (in the Ito convention) 
\begin{align}
	\frac{\partial P}{\partial t}=-\frac{\partial}{\partial x_i}[(A_ix_i^\alpha-\phi(t)x_i)P]+\frac{1}{2}\frac{\partial^2}{\partial x_i^2}(B^2x_i^{2\beta}P)
	\label{eq:fpblumm}
\end{align}
This equation cannot be solved in the general case, but if we neglect fluctuations of $\phi(t)\approx \phi_0$, the time independent steady-state solution $P_0(x_i|A_i)$ of  Eq.~\ref{eq:fpblumm} reads (up to a normalization constant)
\begin{align}
	P_0(x_i|A_i)\propto x_i^{-2\beta}\mathrm{e}^{\frac{2A_i}{B^2}\frac{x_i^\delta}{\delta}
		\left[1-\left(\frac{x_i}{x_c}\right)^{1/\gamma}\right]}
\end{align}
where $\delta=1+\alpha-2\beta$, $\gamma=1/1-\alpha$, and 
\begin{align}
	x_c=\left(\frac{A_i}{\phi_0}\right)
	\left(
	\frac{\delta+1/\gamma}{\delta}
	\right)^\gamma
\end{align}

The most probable value $x_i^*$ for $x_i$ satisfies $dP/dx_i=0$ which implies from Eq.~\ref{eq:fpblumm}
\begin{align}
	F(x_i) = Ax_i^\alpha-\phi_0x_i-B^2\beta x_i^{2\beta-1}=0
\end{align}
For $B=0$, we obtain $x_i^*=0$ or
\begin{align}
	x_i^*=\left(\frac{A_i}{\phi_0}\right)^\gamma
\end{align}
Writing $\sum_ix_i^*=1$ then gives $\phi_0=(\sum_iA_i^\gamma)^{1/\gamma}$. The stability of the solution is given by the sign of $F'(x_i)$ and in this case it is easy to show that the non-zero solution is stable for $A_i>0$ and $\alpha<1$.  

When there is noise ($B\neq 0$), it shifts the deterministic solution $x_i^*$ to a new value. If the noise is not too strong ($B<B_c$), there is a non-zero stable solution and the solution $x_i=0$ which is unstable. In this low  noise case, the score of an item $i$ will be localized around a value  given by the interplay between its fitness and the noise. At $B=B_c$, the non-zero solution disappears and for $B>B_c$, the distribution behaves as $P(x_i|A_i)\sim x_i^{-2\beta}$ and $x_i$ has large fluctuations. In this case, the value of $x_i$ can have values very different from $x_i^*$ and can vary over many orders of magnitude.  

\begin{figure}[ht!]
	\centering
	\includegraphics[width=0.9\linewidth]{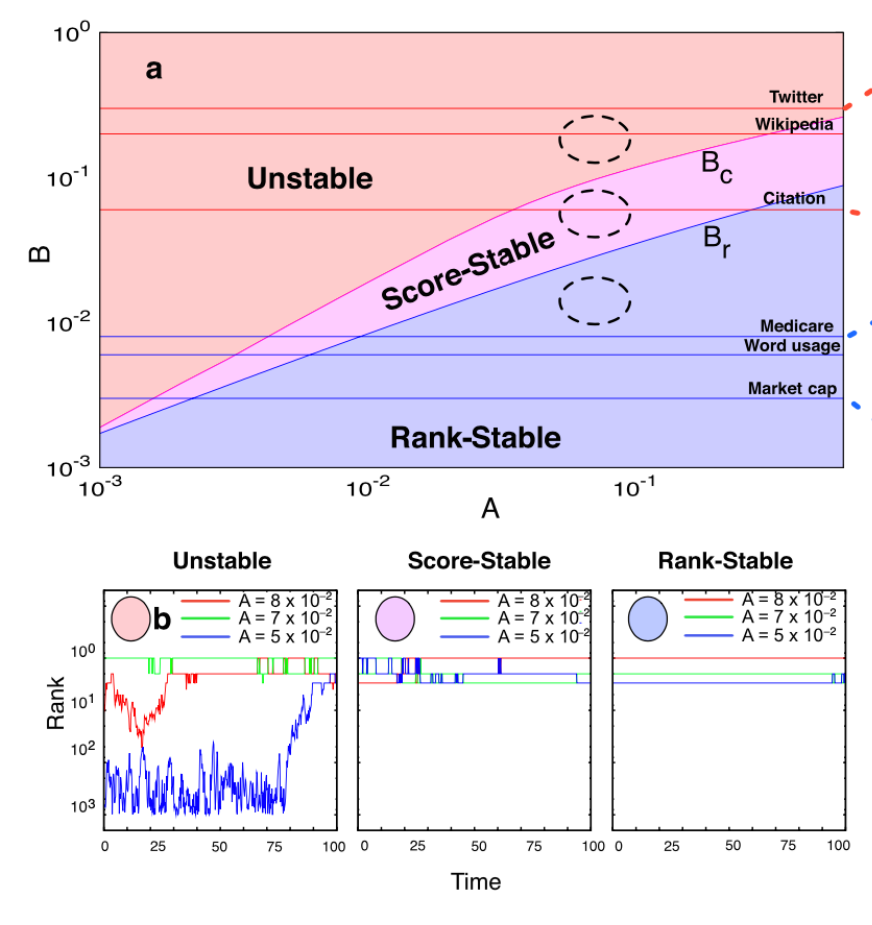}
	\caption{Ranking stability: (a) Phase diagram computed numerically. (b) Time evolution of the ranks for items in each of the three predicted phases (only top-fitness items are shown here). Figure from \protect\cite{Blumm}.}
	\label{fig:blumm2}
\end{figure}

This last discussion is about the value of the score $x_i$, but knowing its value is not enough for determining the rank of item $i$. The rank indeed depends on the values of the other items and as such, is a collective measure: the rank of $i$ depends on the score $x_i$ and also on the scores of all the other items $j$. An item can be `score-stable' with small fluctuations around $x_i^*$ but large enough so that items with similar $x^*$ can  swap ranks. The score and rank stability can occur in the small fluctuations regime $B<B_c$ only, and the rank stability condition reads
\begin{align}
	\langle x\rangle_r-\langle x\rangle_{r+1}>\sigma_r
\end{align} where $\sigma_r$ denotes the rank fluctuations of an item at rank $r$ and $\langle x\rangle_{r(r+1)}$ the average score value at rank $r$ ($r+1$). This condition predicts a second critical value $B_r$ of the noise coefficient. The resulting stability diagram is shown in Fig.~\ref{fig:blumm2}.
The top ($B>B_c$) corresponds to the unstable region where the score is broadly distributed and neither score nor rank stability are possible. In the region $B<B_c$, we observe score stability but as discussed above this is not enough to ensure rank stability. In the region $B_r<B<B_c$ the score is stable but not the rank, each item has a score that fluctuates around its steady state $x^*$ determined by its intrinsic fitness.  For lower noise $B<B_c$, both scores and ranks are stable. These predictions are verified by the temporal profile of ranks shown in Fig.~\ref{fig:blumm2}(b) in each case. 

This simple model thus allows to discuss the variety of behaviors resulting from the interplay between the intrinsic fitness of an item and the noise existing in the system. In the context of cities, we could well imagine that the fitnesses $A_i$ can be very different leading to various simultaneous behavior \cite{Batty:2006}. For large cities, the fitness could be very large and most cities would then be in the rank-stable region. For smaller cities with a smaller fitness, depending on the noise, we could be in different regions of the phase diagram Fig.~\ref{fig:blumm2}(a). Of course, this model is not justified from basic principles and should probably not be trusted in its details, but has the virtue to connect fluctuations of scores and ranks. In this respect, it is a very useful guide for understanding the rank dynamics. 

There is however an important ingredient that this phenomenological model misses: the number of items is not fixed: for cities, new cities can emerge, others can disappear and there is a total net flux of new element in the ranking list. Indeed, ranking lists have typically a fixed size $N_0$ such as the top 100 universities or the Fortune 500 companies etc. The different items can then enter of leave the list at any time of the observation period (denoted here by $t=0,1,\dots,T-2$). For given values of $N_0$ and $T$ it is then possible to introduce two measures of the flux \cite{Iniguez}. In order to understand the dynamics of ranks in various systems, Iniguez et al. \cite{Iniguez} proposed a simplified model that implements only two mechanisms: (i) random displacements of elements across the list, and (ii) random replacements of elements by new ones. A more precise discussion of this problem and modeling in the context of cities  
is missing and would be extremely interesting. More generally, the relation between scores and ranks needs still to be elucidated for general random dynamics.

\section{Integrating interurban migration}

As we saw in the empirical section, the Zipf exponent is not always equal to one, and the rank plot is not always a nice power law. These numerous and important deviations suggest that the Gabaix model proposed to explain this $\nu=1$ behavior might miss some important ingredient. Here, we will follow another line, where we won't try to explain Zipf's law but rather try to construct the evolution equation for urban population starting from first principles.

\subsection{Diffusion with noise: The Bouchaud-Mezard model}

\begin{figure}[!ht]
	\centering
	\includegraphics[width=0.6\linewidth]{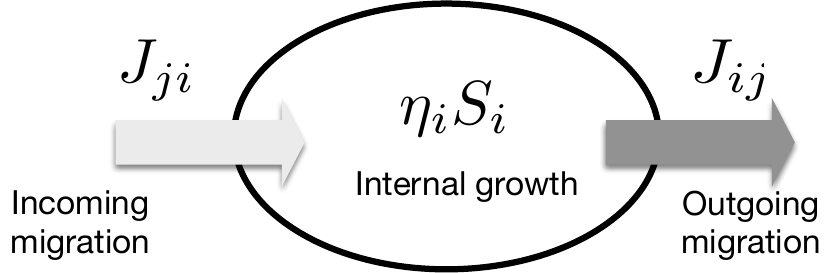}
	\caption{Schematic representation of Eq.~(\ref{eq:stoch}). There is an
		internal growth and `migrations' going either out or in and described
		by the matrix $J_{ij}=I_{ij}W_j$.}
	\label{fig:stoch}
\end{figure}

The `regularization' of the Gibrat model proposed by Gabaix relies on
the assumption of a minimum size of cities which introduces some kind of friction and allows the system to converge to a steady state. We however know from history that cities can shrink, and can even disappear. The assumptions used in the Gabaix model are therefore difficult to defend. A different, simple way to regularize this behavior is to introduce
exchanges between the different constituents of the system (which would then correspond
to migration between different cities). In \cite{BouchaudMezard}, Bouchaud and Mezard
wrote a stochastic dynamical equation in the case of the wealth $W_i(t)$ of an agent $i$ at time $t$
that takes into account trading between individuals. The evolution equation they consider is 
the following  \cite{BouchaudMezard}
\begin{equation}
	\frac{dW_i}{dt}=\eta_i(t)W_i(t)+\sum_jI_{ji}W_j(t)-I_{ij}W_i(t)
	\label{eq:stoch}
\end{equation}
where the first term describes the spontaneous variation of wealth (due to investment for example), and the other term involving the matrix $I_{ij}$ describes the amount of wealth that an agent $j$ spends buying to agent $i$. In the context of cities, the first term is the natural demographic growth, while the second term represents inter-urban migrations (see Fig.~\ref{fig:stoch}).

The random variables $\eta_i(t)$ are assumed to be identically
independent gaussian variables with the same mean $m$ and variance
given by $2\sigma^2$. The flow (per unit time) between agents $i$ and
$j$ is denoted by $J_{ij}=I_{ij}W_j$ and for a general form of these couplings
the solution of the equation \ref{eq:stoch} is unknown. We can however discuss the
simple case of the complete graph where all units are exchanging with
all others at the same rate taken equal to $I_{ij}=I/N$ where $N$
is the number of agents (this scaling ensures a well-behaved large $N$
limit). The equation for $W_i$ becomes \cite{BouchaudMezard}
\begin{equation}
	\frac{dW_i}{dt}=I(\overline{W}-W_i)+\eta_i(t)W_i
	\label{eq:diffnoise}
\end{equation}
where $\overline{W}(t)=\sum_iW_i(t)/N$. We see that the first term act as a homogenizing force
towards the average $\overline{W}$.  All agents feel the same
environment which shows the mean-field nature of this simplified
case. Formally we can treat this equation as an equation for $W_i$
subjected to a source $\overline{W}(t)$ and by integrating formally we
obtain
\begin{equation}
	W_i(t)=W_i(0)\mathrm{e}^{\int_0^t(\eta_i(\tau)-I)\mathrm{d}\tau}+
	I\int_0^t\mathrm{d}\tau\overline{W}(\tau)\mathrm{e}^{\int_\tau^t(\eta_i(\tau')-I)\mathrm{d}\tau'}
\end{equation}
The average quantity $\overline{W}$ is still unknown at this point and
if we sum this equation over $i$ we still can not solve
it. However, if we assume that in the large $N$ limit, the quantity
$\overline{W}$ is self-averaging
\begin{equation}
	\overline{W}\approx_{N\rightarrow\infty}\langle\overline{W}\rangle
\end{equation}
where the brackets $\langle\cdot\rangle$ denote the average over the
variable $\eta$, we have
\begin{equation}
	\overline{W(t)}\simeq\overline{W(0)}\mathrm{e}^{(m+\sigma^2-I)t}+
	I\int_0^t\mathrm{d}\tau\overline{W}(\tau)\mathrm{e}^{(m+\sigma^2-I)(t-\tau)}
\end{equation}
(we assumed here that the initial conditions are independent from the
$\eta$'s). This equation can be easily solved (by Laplace transform for example)
and we obtain
\begin{equation}
	\overline{W(t)}=\overline{W(0)}\mathrm{e}^{(m+\sigma^2)t}
\end{equation}
In order to observe a stationary distribution we normalize $W_i$ by
$\overline{W(t)}$ and construct the variables
$w_i=W_i(t)/\overline{W(t)}$. These quantities obey the following
Langevin-type equation
\begin{equation}
	\frac{dw_i}{dt}=f(w_i)+g(w_i)\delta\eta_i(t)
\end{equation}
where $f(w)=I(1-w)-\sigma^2w$ and $g(w)=w$ (and $\delta\eta=\eta-m$).
In order to write the Fokker-Planck equation for the distribution
$\rho(w)$, we have to give a prescription about the multiplicative
noise. For this we have specify the correlations in the product and
there are basically two main prescriptions, the Ito and Stratonovich
ones. Bouchaud and Mezard used the Stratonovich prescription and obtained the Fokker-Planck
equation for the distribution $\rho(w)$ under the form
\begin{equation}
	\frac{\partial \rho}{\partial t}=-\frac{\partial}{\partial
		w}\left[f\rho\right]+\sigma^2\frac{\partial}{\partial
		w}\left[g\frac{\partial}{\partial w}g\rho\right]
\end{equation}
The equilibrium distribution which satisfies $\partial \rho/\partial t=0$ leads to
\begin{equation}
	\rho(w)=\frac{1}{\cal{N}}\frac{\mathrm{e}^{-\frac{\mu-1}{w}}}{w^{1+\mu}}
\end{equation}
where 
$\cal{N}$  is a normalization constant and where the exponent is given by
\begin{equation}
	\mu=1+\frac{I}{\sigma^2}
	\label{eq:mu}
\end{equation}
This result implies a number of consequences. First, when $I$ is small
(and non zero), we observe that this regularization changes the
lognormal distribution to a power law (at large $w$) with exponent
between $1$ and $2$ (for $I/\sigma^2$ small). This distribution has a
finite average $\langle w\rangle=1$ but an infinite variance,
signaling large fluctuations. In the case where we consider cities (instead of the wealth $W_i$ we then have the population $P_i$ of city $i$), this regularization thus provides a
simple explanation for the diversity of exponents observed in various
countries (see Fig.~\ref{fig:cotti} and \cite{Soo}). The result is
similar to what is obtained with the Gabaix model but with an exponent whose value depends on migration between cities. The
diffusion model discussed here shows that the origin of the Zipf law
lies in the interplay between internal random growth and exchanges
between different cities (we note that this idea that interactions between individual is crucial for understanding Zipf's law was already suggested in \cite{Marsili}). An important consequence is that increasing mobility should actually increase $\mu$ and therefore reduces the
heterogeneity of the city size distribution. Few words of caution are
however needed here: first the relation Eq.~(\ref{eq:mu}) should be
tested empirically, and second, the important assumption that $I_{ij}$
is constant is by no means obvious.

\subsection{From first principles to the growth equation of cities}

The Bouchaud-Mezard equation \ref{eq:stoch} tells us that the distribution results from the interplay between random growth and exchanges between the units of the system. This leads to the idea that inter-urban migration plays a crucial role in shaping the population distribution. Instead of trying to explain Zipf's law, we then proceed differently and start from first principles \cite{Verbavatz}. More specifically, we can enumerate all the possible sources of population variations. The dynamics of a system of cities (such as a country) can be decomposed into a sum of various terms, that we write under the following form
\begin{align}
	\partial_t P_i=F_{demo}+F_{internat}+F_{interurb}
\end{align}
This expresses the fact that the variation of the population has three main sources: the first one $F_{demo}$ is the demographic growth which captures both births and deaths and that we can write under the form $F_{demo}=\eta_i P_i$ where $\eta_i$ is the demographic growth rate of city $i$. The second term $F_{internat}$ describes international migrations (individuals coming from other countries to the city $i$). This term is usually small and we will integrate it in the growth rate $\eta_i$. The last term $F_{interurb}$ describes interurban migrations, ie. individuals moving from a city to another and can be written as 
\begin{align} 
F_{interub}=\sum_{j\in N(i)}[J_{j\to i}-J_{i\to j}]
\end{align} 
 where $N(i)$ the set of neighbors of city $i$ (ie. cities that exchange a non-zero number of inhabitants), and where $J_{j\to i}$ is the migration rate from city $j$ to city $i$. The time evolution of the population size $P_i$ can then be written as
\begin{align}
	\partial_t P_i=\eta_i P_i+\sum_{j\in N(i)}[J_{j\to i}-J_{i\to j}]
	\label{eq:general}
\end{align}
We note that if there is an exact balance of migration flows $J_{i\to j}=J_{j\to i}$, the equation becomes equivalent to Gibrat’s model discussed above and which predicts a lognormal distribution of populations. 

\begin{figure}[!ht]
	\centering
	\includegraphics[angle=-90,scale=0.3]{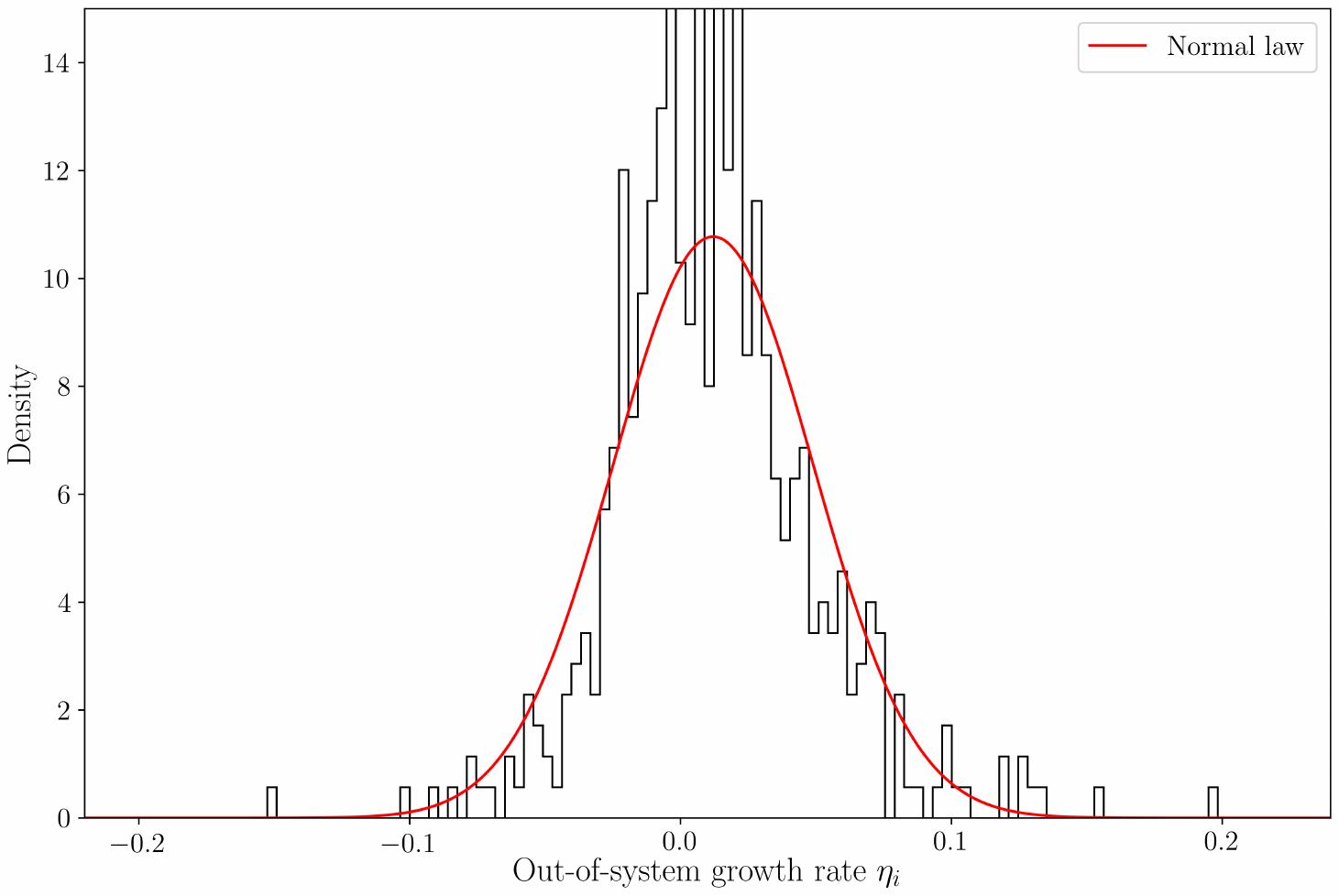}
	\caption{Distribution of the growth rate due to natural growth, international migrations, exchanges with small towns. The data shown is for the French cities in 2003-2008 and compared to a normal distribution. Figure from \cite{Verbavatz}.}
	\label{fig:Peta}
\end{figure}

We started from general considerations (as it amounts to write the balance of births, deaths and migrations), and ended up with this general  stochastic equation \ref{eq:general} which is the diffusion with noise equation discussed in the previous section. This equation is very general, and we have to specify the various terms. In particular the main difficulty is to find a reasonable simplified expression for the migration terms. The idea is then to characterize these different terms empirically. In \cite{Verbavatz}, we focused on four recent datasets of migrations  (USA for 2012-2017, France for 2003-2008, UK for 2012-2016 and Canada for 2012-2016)  and in  \cite{Satish}, measures were redone on other datasets confirming these results. We found that the quantity $\eta_i$ is an uncorrelated random variable distributed according to a Gaussian with average $r$ (the average growth rate) and variance $\sigma$ (see Fig.~\ref{fig:Peta}). 

\begin{figure}[!ht]
	\centering
	\includegraphics[scale=0.3]{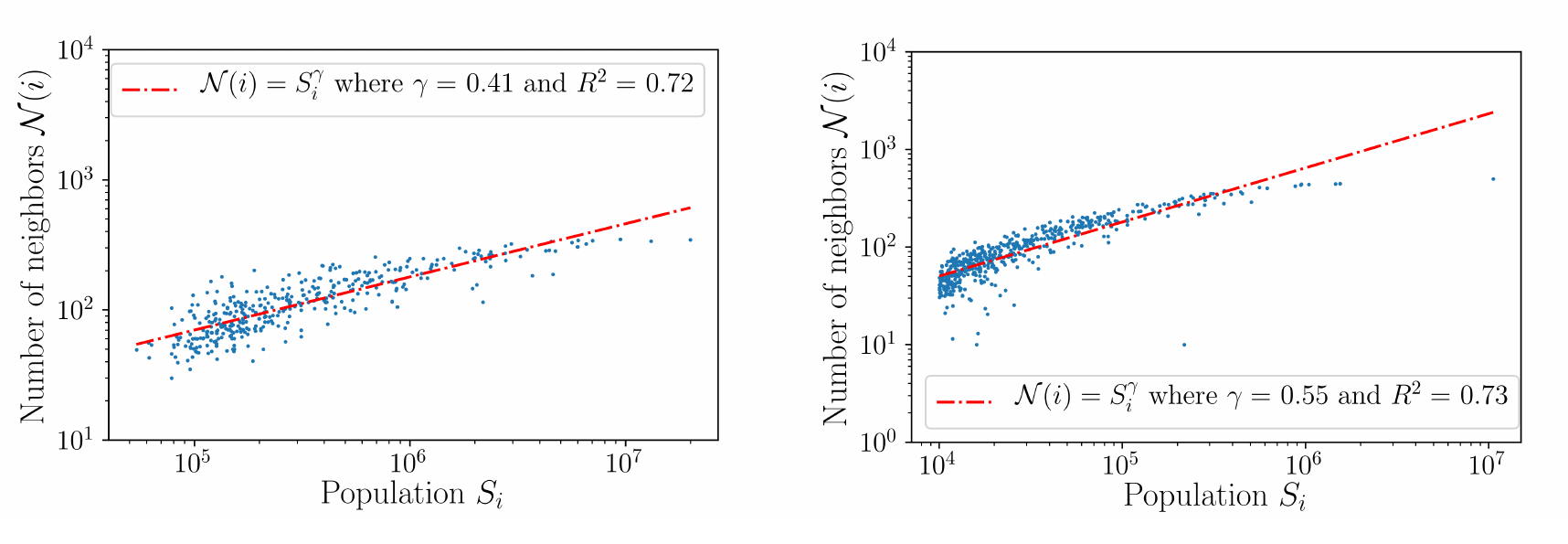}	
	\caption{Number of neighbors for each city as a function of population for the US (a) and France (b). The dotted red lines indicate the power law fit $N(i) \sim P_i^\gamma$. In the UK and in Canada, we have a fully-connected dataset and $\gamma = 0$. Figure from \cite{Verbavatz}.}
	\label{fig:Ni}
\end{figure}

We also found in these datasets that for France and the US, the number of neighbors exchanging a non-zero number of individuals is $N(i)\sim P_i^\gamma$ where $\gamma\approx 0.4-0.5$. For the UK and Canadian datasets (that are smaller), all pairs of cities exchange individuals between each other and we get $\gamma=0$ (see Fig.~\ref{fig:Ni}).

The important and difficult term is the interurban migration sum. The migration flow $J_{i\to j}$ depends a priori at least on the populations $P_i$ and $P_j$, and the distance $d_{ij}$ between cities $i$ and $j$. Using the standard so-called `gravitational' model of the form \cite{Erlander}
\begin{align}
	J_{i\to j}=K\frac{P_i^\mu P_j^\nu}{d_{ij}^\sigma}
\end{align}
we showed that the dominant contribution to the migration flows comes from the populations and that the distance is a second-order effect.  This is actually reasonable as distance is not a strong barrier when it comes to move from a location to a new one.  This suggests that we can write the migration term under the following form \cite{Verbavatz}
\begin{align}
	J_{i\to j}=I_0 P_i^{\nu '} P_j^\nu x_{ij}
\end{align}
where the random variables $x_{ij}$ have an average equal to $1$ and encode all the noise and multiple effects, including distance. The data also shows that we have $\nu'=\nu$ and that on average there is a sort of detailed balance with $J_{i\to j}=J_{j\to i}$, but also and crucially, that there are fluctuations that can be large. 
More precisely, if we denote by 
\begin{align}
	X_{ij}=[J_{i\to j}-J_{j\to i}]/(I_0 P_i^\nu)
\end{align}
these random variables are distributed according to a broad law with heavy tails that decay as power law with exponent $\alpha <2$.  
 \begin{figure}[!h]
	\centering
	\includegraphics[scale=0.3]{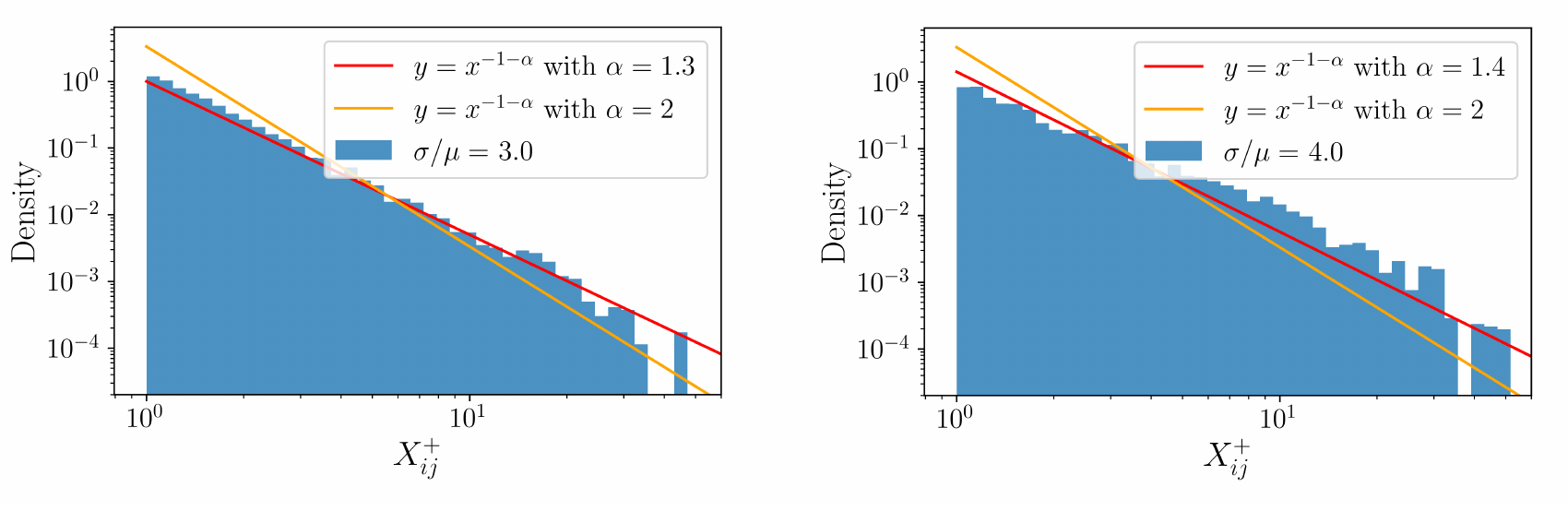}
	\caption{Density of the right-tail quantity $X+$ (ie. $X_{ij}>0$) for the US (left) and France (right). Both distributions have heavy tails and display a relative dispersion $\sigma/m = 3.0$ and $4.0$ for the US and France respectively. The tail of both distributions is asymptotically described by a power law with exponent $1+\alpha$ where $\alpha<2$. Figure from \cite{Verbavatz}.}
	\label{fig:PX}
\end{figure}

The sum in the second term of the r.h.s. of Eq.~\ref{eq:general} can then be rewritten as 
\begin{align}
	\sum_{j \in N(i)}J_{j\to i}-J_{i\to j}=I_0P_i^\nu\sum_{j\in N(i)}X_{ij}
\end{align}
This expression is a sum of random variables and its distribution can be found using central limit theorems \cite{Gnedenko}. There is however a subtlety here: the random variables are broadly distributed, and have an infinite second moment. The usual central limit theorem cannot be applied in this case and the limiting distribution is not the usual gaussian law. Instead, we have to use the generalized version of the central limit theorem \cite{Gnedenko} (and assuming that the correlations between the variables $X_{ij}$ are negligible), that shows that the random variable
\begin{align}
	\zeta_i=\frac{1}{N(i)^{1/\alpha}}\sum_{j\in N(i)} X_{ij}
\end{align}
follows (for a large enough N(i)) a L\'evy stable law $L_\alpha$ of parameter $\alpha$ and scale parameter $s$.

Putting all these elements together, we are led to the conclusion that the growth of systems of cities is governed by a stochastic differential equation of a new type with two independent noises:
\begin{align}
	\partial_t P_i=\eta_i P_i+DP_i^\beta\zeta_i
\label{eq:oureq}
\end{align}
where $D=sI_0$, $\beta=\nu+\gamma/\alpha$ and where $\eta_i$ is a gaussian noise of mean  the average growth rate $r$ and dispersion $\sigma$. Empirically, we observe that $\beta<1$ for all cases.

Eq. \ref{eq:oureq} is the stochastic equation of cities that governs the dynamics of large urban populations and which is our main result here. In this equation both noises $\eta$ and $\zeta$ are uncorrelated, multiplicative and Ito’s convention \cite{VanKampen} seems here to be the more appropriate as population sizes at time $t$ are computed independently from inter-urban migrations terms at time $t+dt$. 

The central limit theorem together with the broadness of inter-urban migration flow allowed us to show that many details in Eq. \ref{eq:general} are in fact irrelevant and that the dynamics can be described by the more universal Eq. \ref{eq:oureq}. Starting from the exact equation \ref{eq:general} is therefore not only doomed to failure in general, but is also not useful. The importance of migrations was already sensed in \cite{Bettencourt}, but the authors derived a stochastic differential equation with multiplicative Gaussian noise, which we show here to be incorrect: we have indeed a first multiplicative noise term but also crucially another term that is a multiplicative L\'evy noise with zero average. This is a major and non-trivial theoretical shift that was missed in all previous studies on urban growth and which has many capital implications, both in understanding stationary and dynamic properties of cities.

\subsubsection{Distribution}

\begin{figure}[!ht]
	\centering
	\includegraphics[scale=0.5]{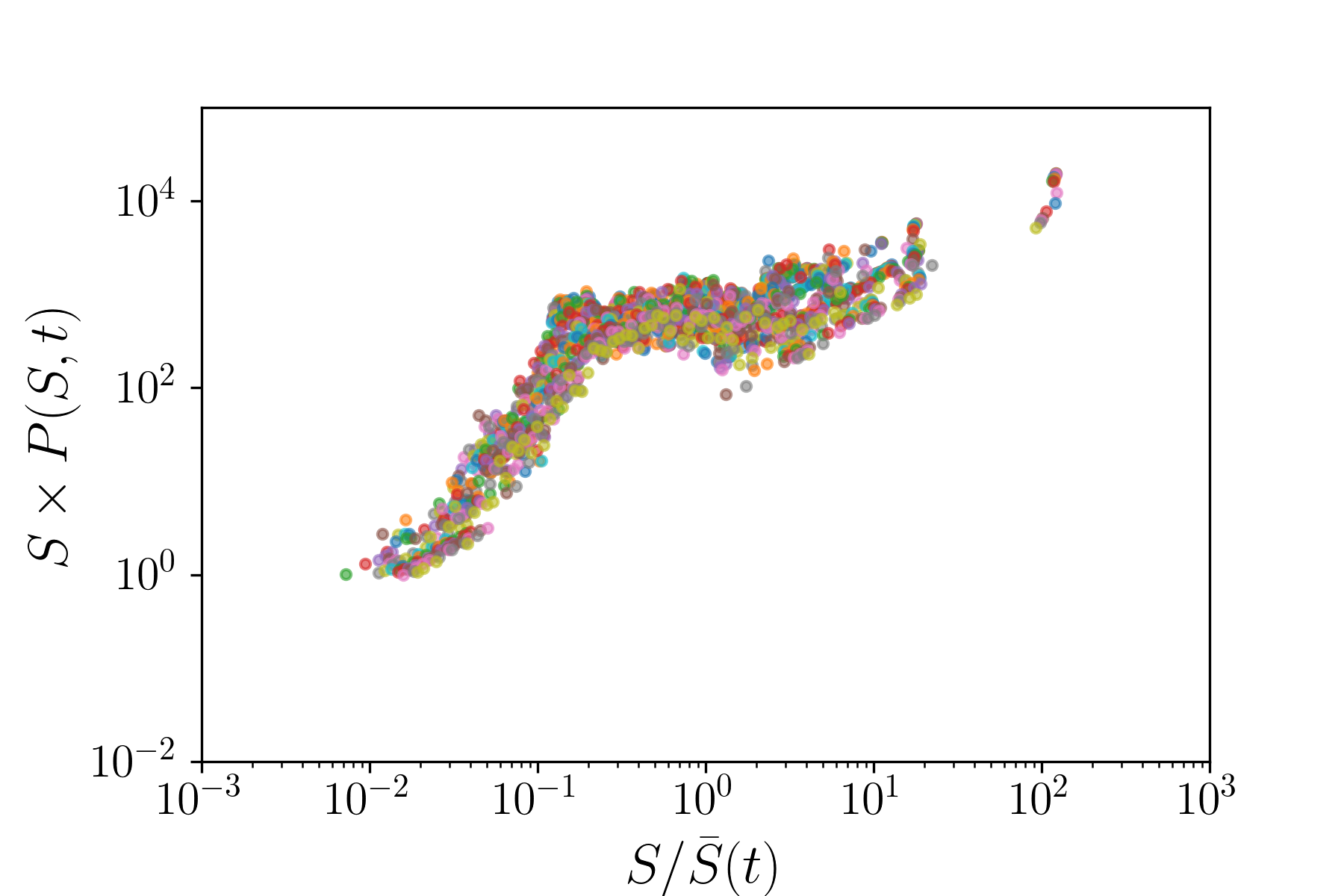}
	\caption{Scatterplot of the quantity $\rho(P, t) P$ versus the ratio $P/\overline{P}(t)$ for France's largest $500$ cities between 1875 and 2016. Each color is a different year. We observe that the plots of all years collapse towards a unique universal function of $P/\overline{P}(t)$ in agreement with the result of Eq.~\ref{eq:colla}. Figure from \cite{Verbavatz}.}
	\label{fig:colla}
\end{figure}

The equation \eqref{eq:oureq} governs the evolution of urban population and analyzing it at large times gives indications about the stationary distribution of cities. In order to discuss analytical properties of this equation, we assume that Gaussian fluctuations are negligible compared to the L\'evy noise and write $\eta_i\approx r$. The corresponding Fokker-Planck equation (with Ito’s convention) can be solved approximately in the limit of large populations using the formalism of Fractional Order Derivatives and Fox functions \cite{Srokowski,Jespersen,Schertzer,Fox}. We then obtain the general distribution at time $t$ that can be expanded in powers of $P$ as \cite{Srokowski}:
\begin{align}
\rho(P,t)=\sum_{k=1}^{\infty}C_k\left(\frac{1}{a(t)P}\right)^{\alpha\beta+\alpha(1-\beta)k}
\label{eq:expan}
\end{align}
where $C_k$ is a complicated prefactor independent from time and $P$ and where 
$a(t)$ decreases exponentially at large times.  Eq. \ref{eq:expan} shows that there is a scaling law of the form
\begin{align}
	\rho(P,t)=\frac{1}{P} F\left(\frac{P}{\overline{P}(t)}\right)
	\label{eq:colla}
\end{align}
with a scaling function $F$ that depends on the country only. We confirmed this scaling form for France (the only country for which we had sufficient data) and the resulting collapse is shown in Fig.~\ref{fig:colla}. This good collapse confirms the validity of the expansion Eq.~\ref{eq:expan}.

\begin{figure}[!ht]
	\centering
	\includegraphics[scale=0.5]{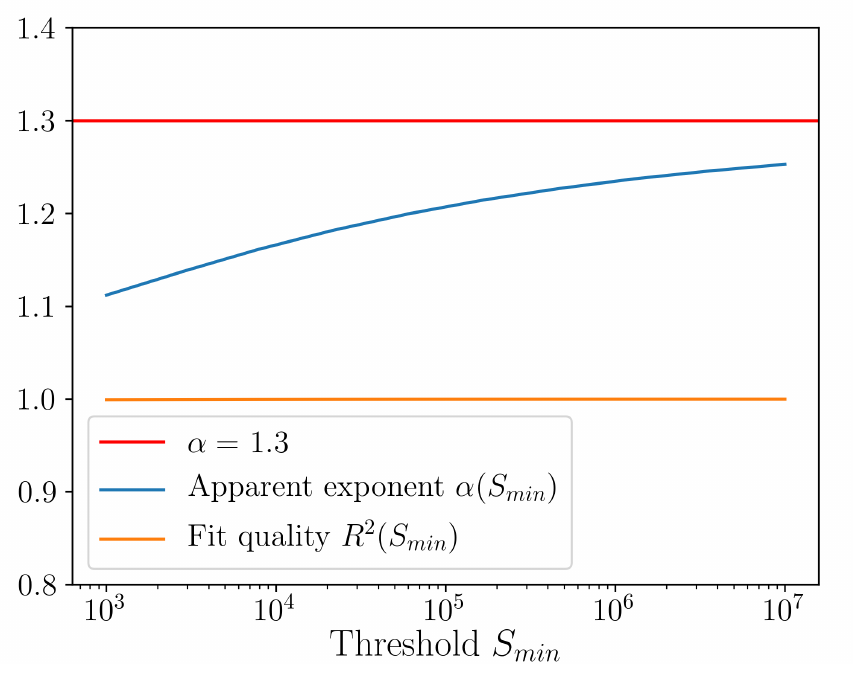}
	\caption{Power law fit of the expansion Eq.~\ref{eq:expan}. This fit gives an apparent exponent $\alpha(S_{\min}$ with very good quality ($R^2\approx 1$), although the expansion itself is not a power law. The apparent exponent is smaller than $\alpha$ (chosen here equal to $\alpha=1.3$), but slowly converges towards $1.3$ as the value of the threshold $S_{\min}$ increases (parameters are here $\alpha = 1.3$, $\beta = 0.8$, $r = 0.01$, $D = 0.06$ and $t = 500$). Figure from \cite{Verbavatz}.}
	\label{fig:effective}
\end{figure}

This expansion Eq. \ref{eq:expan} allows us to discuss the validity of Zipf's law in general.  Indeed, we observe that the population distribution is dominated at large $P$ by the order $k=1$ and converges towards a power law with exponent $\alpha \neq 1$. The speed of convergence towards this power-law can be estimated with the ratio $\lambda(P,t)$ of the second  over the first term of the expansion Eq. \ref{eq:expan} and reads
\begin{align}
	\lambda (P,t)=\frac{D^{\alpha}}{r}(\overline{P(t)}/P)^{\alpha (1-\beta)}
\end{align}
where $\overline{P(t)}$ is the average city size. If $\lambda (P)\geq 1$, the power law regime with exponent $\alpha$ is not dominant (and cannot be directly observed). Estimates of $\alpha$ and $\beta$ for the four datasets show that finite-time effects are very important in general and that a power-law regime is only reached for unrealistically large city sizes. For most cases, there is therefore no reason to observe a clear power law with a universal exponent, and a fortiori a universal Zipf's law of the form $P\sim 1/r^\nu$. We can however explain why empirically such a Zipf's law for cities was apparently measured in many cases. If we perform a power law fit directly on the solution Eq.~\ref{eq:expan}, we obtain an `effective' exponent value that depends on many details (such as the range over which the fit is made). We show on Fig.~\ref{fig:effective} an example of such a fit over a range $[S_{\min},S_{\max}]$.
The possibility to have a good power law fit might explain why an apparent Zipf's law was observed for such a long time, and also why it is not universal. The effective exponent depends on many details and this might explain the large diversity of exponents collected in \cite{Cottineau}. 
 
\subsubsection{Dynamics}
 
When we observe these historical developments about Zipf's law, it is natural to think that it is not enough for an equation to explain the population distribution for demonstrating its validity. Further tests are needed, and in particular such an equation as Eq.~\ref{eq:oureq}, if correct, should also give information about the dynamics of cities over time. This can be done by following the populations and ranks of the system’s cities at different times with the help of rank clocks introduced in \cite{Batty:2006}. In this work, it was proven that the micro-dynamics of cities is very turbulent with many rises and falls of entire cities that cannot result from Gabaix’s model.

\begin{figure*}[!ht]
\centering
\includegraphics[scale=0.45]{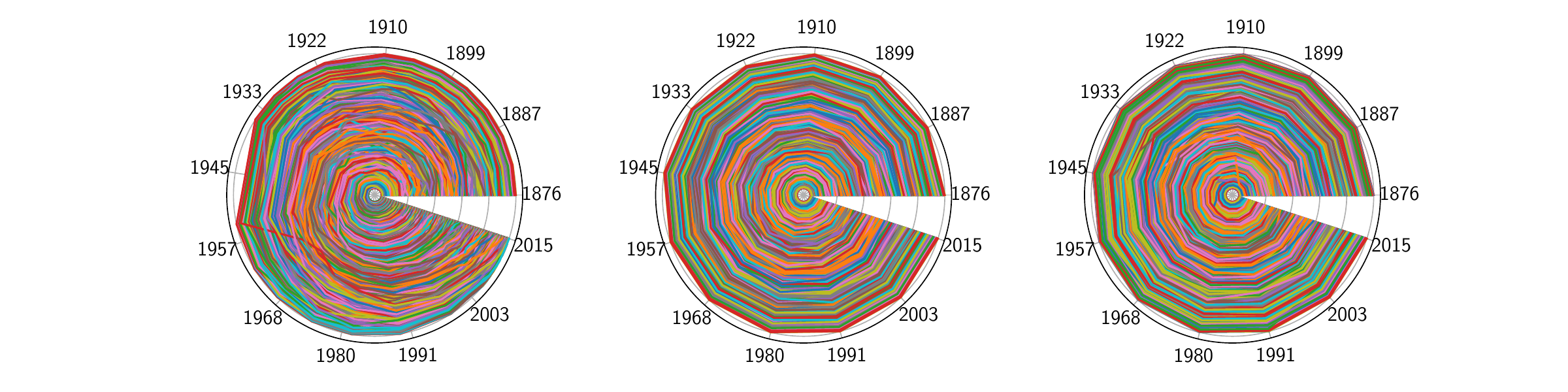}
\caption{Rank clocks for France. We compare the real dynamics of the $500$ largest French cities between 1876 and 2015 (left) to Gabaix’s statistical prediction (middle) and to the model Eq.~\ref{eq:oureq} (right). On the clocks, each line represents a city rank over time where the radius is given by the rank and the angle by time. In this representation, the largest city is at the center and the smallest at the edge of the disk. Figure from \cite{Verbavatz}.}
\label{fig:rankclock}
\end{figure*}

\begin{figure}[!ht]
	\centering
	\includegraphics[scale=0.5]{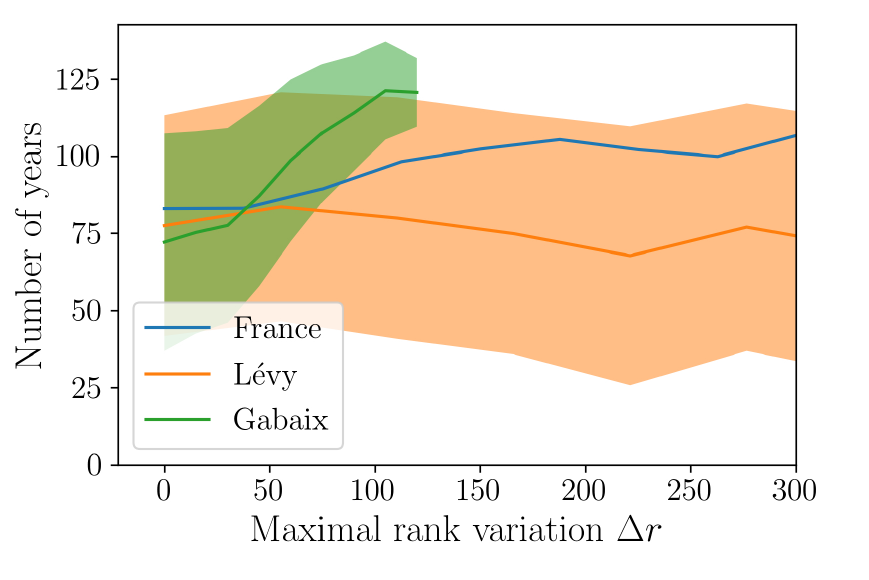}
	\caption{Average number of years taken to observe the maximal rank variation $\Delta r$. Although dispersion is large, L\'evy’s model is compatible with real data in contrast to Gabaix’s model of growth. Figure from \cite{Verbavatz}.}
	\label{fig:number_years}
\end{figure}

In order to compare various models, we show in Fig.~\ref{fig:rankclock} the empirical rank clock for France (from 1876 to 2015) and for numerical results obtained with Gabaix’s model and Eq.~\ref{eq:oureq}. 
We see that in Gabaix’s model Fig.~\ref{fig:rankclock}(middle), city ranks are on average stable and not turbulent: the rank trajectories are concentric and the rank of a city oscillates around its average position. In the real dynamics (left), cities can emerge or die. Very fast changes in rank order can occur, leading to a much more turbulent behavior. In the model described by Eq.~\ref{eq:oureq}, the large fluctuations of L\'evy’s noise are able to statistically reproduce such fast surges and swoops of cities  (Fig.~\ref{fig:rankclock}(right)). We can quantity this a bit more precisely \cite{Verbavatz}, For example, we compare the average shift per time
\begin{align}
	d=\frac{1}{NT}\sum_{t}\sum_{i=1}^N r_i(t)-r_i(t-1)
\end{align}
over $T$ years and for $N$ cities in the three cases, or study the statistical fluctuations of the rank. For all these measures, we find that L\'evy fluctuations are much more able to reproduce the turbulent properties of the dynamics of cities through time. Indeed, the fast births or deaths of cities due for example to wars, discoveries of new resources, incentive settlement policies, etc. are statistically explained by broadly-distributed migrations and incompatible with a Gaussian noise. We can also compare with the empirical data the predictions of the different models for the time needed to make the largest rank jump (see Fig.~\ref{fig:number_years}).
We observe that for France, there is typically a duration of order 80 years to make a very large rank jump. We confirm here that Gabaix’s model is unable to reproduce these very large fluctuations and that equation \ref{eq:oureq} agrees very well with the data.

\section{Discussion and perspectives}

We reviewed different approaches based on stochastics differential equations for describing the evolution of urban populations (a more extensive discussion can be found in \cite{BV}). This problem has a long history and showcases some of the danger and pitfalls of stylized facts but also the importance of extensive datasets. Indeed, after the empirical observations of Auerbach and Zipf, most of the theoretical approaches focused on the problem how to explain a Zipf exponent equal to one.  Gibrat's model couldn't explain the appearance of a stationary distribution, implying the necessity of introducing some friction. This is what the Gabaix model did with a simple mechanisms that produces indeed a Zipf exponent equal to one (or equivalently a power law distribution for the population with exponent equal to $2$). Other models were then proposed and specifically designed to produce a Zipf exponent equal to one. However the recent availability of massive datasets, for many countries and many time periods showed that the Zipf exponent is not universal and could be very different from one country to another. This is what actually forced us to reconsider this problem but with another perspective: instead of trying to explain the Zipf result, we tried to construct the evolution equation starting from basic principles. This led to a stochastic equation of growth for cities that is empirically sound and challenges Zipf's law and current models of urban growth. This equation \ref{eq:oureq} of a new type with two sources of noise predicts an asymptotic power-law regime, but this stationary regime is not reached in general and finite-time effects cannot be discarded. This implies that in general, the Zipf plot is not a power law and a power law fit gives an effective exponent that depends on many details of the system and can vary a lot from a case to another. In other words, Zipf's law for cities does not hold in general. In addition, and in contrast with most existing models, this equation is also able to statistically reproduce the turbulent micro-dynamics of cities with fast births and deaths.

In the derivation of this equation, we showed that microscopic details of interurban migrations are irrelevant and the growth equation obtained is universal. A crucial point in this reasoning is that although we have on average some sort of detailed balance that would lead to a Gaussian multiplicative growth process, it is the existence of non-universal and broadly distributed fluctuations of the microscopic migration flows between cities that govern the statistics of city populations. This result exhibits an interesting connection between the behavior of complex systems and non-equilibrium statistical physics for which microscopic currents and the violation of detailed balance seem to be the rule rather than the exception \cite{Bouchaud}. 

At a practical level, this result also highlights the critical effect of not only inter-urban migration flows (an ingredient that is generally not considered in urban planning theories), but more importantly their large fluctuations, ultimately connected to the capacity of a city to attract a large number of new citizens. From the urban science perspective, the next frontier is to describe the growth of cities, not from the population point of view only, but in spatial terms. The surface growth equation for cities is still missing and the recent availability of remote sensing data for many cities will certainly trigger interesting results in the future. 

There are many other questions that are left open. In particular, the spatial structure of these flows displays interesting features for the US\cite{Satish} and it would be interesting to investigate many countries and different periods. Also, we  discussed a phenomenological approach to the dynamics of ranking. This is certainly an interesting research direction. In particular, if the dynamics of scores of items are governed by a given equation, we generally don't know what it implies for the ranking dynamics. In other words, the mapping from the equation for scores to the equation for ranks is in general missing.

\end{document}